\documentclass{iscram}
\usepackage{graphicx}
\usepackage{color}
\usepackage{url}
\bibliography{bibliography}

\usepackage{tikz}
\usetikzlibrary{plotmarks}

\iscramset{
title={Automatic Image Filtering on Social Networks Using Deep Learning and Perceptual Hashing During Crises},
short title={Automatic Image Filtering on Social Networks},
author={full name={Dat T. Nguyen}, short name={D. T. Nguyen},
affiliation={Qatar Computing Research Institute\\
Hamad Bin Khalifa University\\
Doha, Qatar\\
ndat@hbku.edu.qa}},
author={full name={Firoj Alam}, short name={F. Alam},
affiliation={Qatar Computing Research Institute\\
Hamad Bin Khalifa University\\
Doha, Qatar\\
falam@hbku.edu.qa}},
author={full name={Ferda Ofli}, short name={F. Ofli},
affiliation={Qatar Computing Research Institute\\
Hamad Bin Khalifa University\\
Doha, Qatar\\
fofli@hbku.edu.qa}},
author={full name={Muhammad Imran}, short name={M. Imran},
affiliation={Qatar Computing Research Institute\\
Hamad Bin Khalifa University\\
Doha, Qatar\\
mimran@hbku.edu.qa}},
iscram 2017 footer,
CoRe Paper 2017={Social Media Studies}
}

\begin{document}
\maketitle
\abstract{The extensive use of social media platforms, especially during disasters, creates unique opportunities for humanitarian organizations to gain situational awareness and launch relief operations accordingly. In addition to the textual content, people post overwhelming amounts of imagery data on social networks within minutes of a disaster hit. Studies point to the importance of this online imagery content for emergency response. Despite recent advances in the computer vision field, automatic processing of the crisis-related social media imagery data remains a challenging task. It is because a majority of which consists of redundant and irrelevant content. In this paper, we present an image processing pipeline that comprises de-duplication and relevancy filtering mechanisms to collect and filter social media image content in real-time during a crisis event. Results obtained from extensive experiments on real-world crisis datasets demonstrate the significance of the proposed pipeline for optimal utilization of both human and machine computing resources.
} 

\keywords{social media, image processing, supervised classification, disaster management}



\section{Introduction}
The use of social media platforms such as Twitter and Facebook at the times of natural or man-made disasters has increased recently~\parencite{starbird2010chatter,hughes2009twitter}. People post a variety of content such as textual messages, images, and videos online~\parencite{chen2013understanding,imran2015processing}. Studies show the significance and usefulness of this online information for humanitarian organizations struggling with disaster response and management~\parencite{petersinvestigating,imran2013extracting,daly2016mining}.
A majority of these studies have however been relying almost exclusively on textual content (i.e., posts, messages, tweets, etc.) for crisis response and management. Contrary (or complementary) to the existing literature on using textual social media content for crisis management, this work focuses on leveraging the visual social media content (i.e., images) to show humanitarian organizations its utility for disaster response and management. While many are valuable, however, the sheer amount of images (most of which are irrelevant or redundant) is difficult to manage and digest for humanitarian organizations, let alone useful for emergency management. If processed timely and effectively, this information can enable early decision-making and other humanitarian tasks such as gaining situational awareness, e.g., through summarization~\parencite{rudra2016summarizing} during an on-going event or assessing the severity of the damage during a disaster~\parencite{ofli2016combining}.

Analyzing the large volume of online social media images generated after a major disaster remains to be a challenging task in contrast to the ease of acquiring them from various social media platforms. A most popular solution is to use a hybrid crowdsourcing and machine learning approach to rapidly process large volumes of imagery data for disaster response in a time-sensitive manner. In this case, human workers (paid or volunteers~\parencite{reuter2015xhelp}) are employed to label features of interest (e.g., damaged shelters or blocked roads) in a set of images.
These human-annotated images are then used to train supervised machine learning models to recognize such features in new unseen images automatically.

However, a large proportion of social media image data consists of irrelevant (see Figure~\ref{fig:sample_irrelevant}) or redundant (see Figure~\ref{fig:sample_duplicate}) content. Many people just re-tweet an existing tweet or post irrelevant images, advertisement, even porn using event-specific hashtags. On the other hand, the time and motivation of human annotators are neither infinite nor free. Every crowdsourcing deployment imposes a cost in humanitarian organizations' volunteer base and budget. Annotators may burn down (reducing their effectiveness due to lack of motivation, tiredness, or stress) or drop out completely (reducing the humanitarian organizations' volunteer base). Since human annotations have a direct effect on the performance of the machine learning algorithms, deficiencies in the annotations can easily translate to shortcomings in the developed automatic classification systems. Therefore, it is of utmost importance to have many volunteers to provide annotations (i.e., \emph{quantity}) and mechanisms to keep the annotation \emph{quality} high.

One way to achieve this is to decrease the workload on the human annotators. For this purpose we develop an image processing pipeline based on deep learning that can automatically (i) detect and filter out images that are not relevant or do not convey significant information for crisis response and management, and (ii) eliminate duplicate or near-duplicate images that do not provide additional information neither to a classification algorithm nor to humanitarian organizations. Such a filtering pipeline will thus help annotators to focus their time and effort on making sense of useful image content only, which in turn helps improve the performance of the machine learning models. Among several potential use cases of the proposed image filtering pipeline on social networks, in this particular study, we focus on the use case of automatic damage assessment from images.


The main contributions of this study can be summarized as follows: (i) We propose mechanisms to purify the noisy social media image data by removing duplicate, near-duplicate, and irrelevant image content. (ii) We use the proposed mechanisms to demonstrate that a big chunk of the real-world crisis datasets obtained from online social networks consists of redundant or irrelevant content. (iii) Our extensive experimental evaluations underline the importance of the proposed image filtering mechanisms for optimal utilization of both human and machine resources. Specifically, our experimental results show that purification of social media image content enables efficient use of the limited human annotation budget during a crisis event, and improves both robustness and quality of the machine learning models' outputs used by the humanitarian organizations. (iv) We show that the state-of-the-art computer vision deep learning models can be adapted successfully to image relevancy and damage category classification problems on real-world crisis datasets. (v) Finally, we develop a real-time system with an image processing pipeline in place for analyzing social media data at the onset of any emergency event, which can be accessed at~\url{http://aidr.qcri.org/}.


\section{Related work}



Despite the wealth of text-based analyses of social media data for crisis response and management, there are only a few studies analyzing the social media image content shared during crisis events.
The importance of social media images for disaster management has been recently highlighted in~\parencite{petersinvestigating}.
The authors analyzed tweets and messages from Flicker and Instagram for the flood event in Saxony in 2013, and found that the existence of images within on-topic messages are more relevant to the disaster event, and the image content can also provide important information related to the event.
In another study, \citeauthor{daly2016mining} focused on classifying images extracted from social media data, i.e, Flickr, and analyzed whether a fire event occurred at a particular time and place~\parencite{daly2016mining}. Their study also analyzed spatio-temporal meta-data associated with the images, and suggested that geotags prove useful to locate the fire affected area.

Taking a step further, \citeauthor{chen2013understanding} studied the association between tweets and images, and their use in classifying visually-relevant and irrelevant tweets~\parencite{chen2013understanding}. They designed classifiers by combining features from text, images and socially relevant contextual features (e.g., posting time, follower ratio, the number of comments, re-tweets), and reported an F1-score of 70.5\% in a binary classification task, which is 5.7\% higher than the text-only classification.

There are similar studies in other research domains (e.g., remote sensing), not necessarily using social media data though, that try to assess level of damage from aerial~\parencite{TurkerM:IJRS04,FernandezGalarreta:2015dn} and satellite \parencite{pesaresi2007rapid,feng2014application} images collected from disaster-hit regions. 


Almost all of the aforementioned studies rely on the classical bag-of-visual-words-type features extracted from images to build the desired classifiers. However, since the introduction of Convolutional Neural Networks (CNN) for image classification in \parencite{NIPS2012_4824}, state-of-the-art performances for many tasks today in computer vision domain are achieved by methods that employ different CNN architectures on large labeled image collections such as PASCAL VOC~\parencite{Everingham10} or ImageNet~\parencite{ILSVRC15}. Recently, in the 2016 ImageNet Large Scale Visual Recognition Challenge (ILSVRC)\footnote{http://image-net.org/challenges/LSVRC/2016/results\#loc}, the best performance on image classification task is reported as $2.99\%$ top-5 classification error by an ensemble method based on existing CNN architectures such as Inception Networks~\parencite{SzegedyVISW15}, Residual Networks~\parencite{He_2016_CVPR} and Wide Residual Networks~\parencite{ZagoruykoK16}.

More importantly, many follow-up studies have shown that the features learned automatically by these deep neural networks are transferable to different target domains~\parencite{JDonahue:ICML14,Sermanet:ICLR14,Zeiler:ECCV14,RGirshick:CVPR14,MOquab:CVPR14}. This proves extremely useful for training a large network without overfitting when the target dataset is significantly smaller than the base dataset, as in our case, and yet achieving state-of-the-art performance in the target domain. Therefore, we consider this transfer learning approach in our study.

\section{Data Collection and Annotation}
\label{sec:datasets}
\begin{figure*}[ht!]
\centering
\begin{tabular}{p{0.1cm} p{0.1cm} c c c}
  			& & \textbf{Severe damage} & \textbf{Mild damage} & \textbf{None}  \\
                        
\raisebox{3.2\normalbaselineskip}[0pt][0pt]{\rotatebox[origin=c]{90}{\textbf{Nepal}}}
&\raisebox{3.2\normalbaselineskip}[0pt][0pt]{\rotatebox[origin=c]{90}{\textbf{Earthquake}}}
&\includegraphics[width=0.28\textwidth]{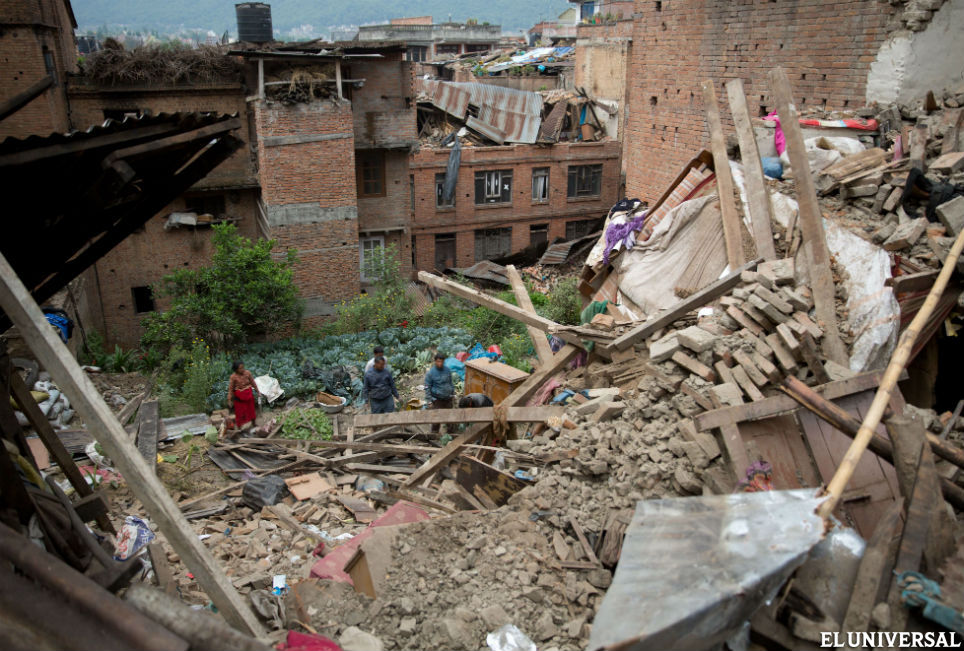}
&\includegraphics[width=0.28\textwidth]{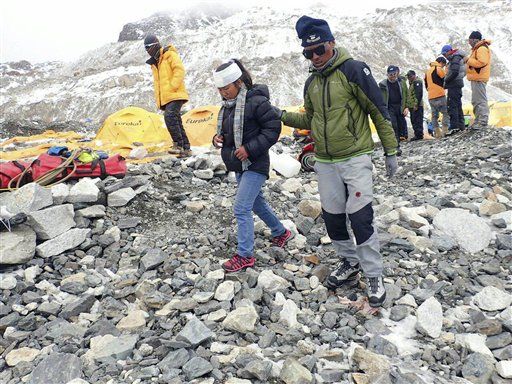}
&\includegraphics[width=0.28\textwidth]{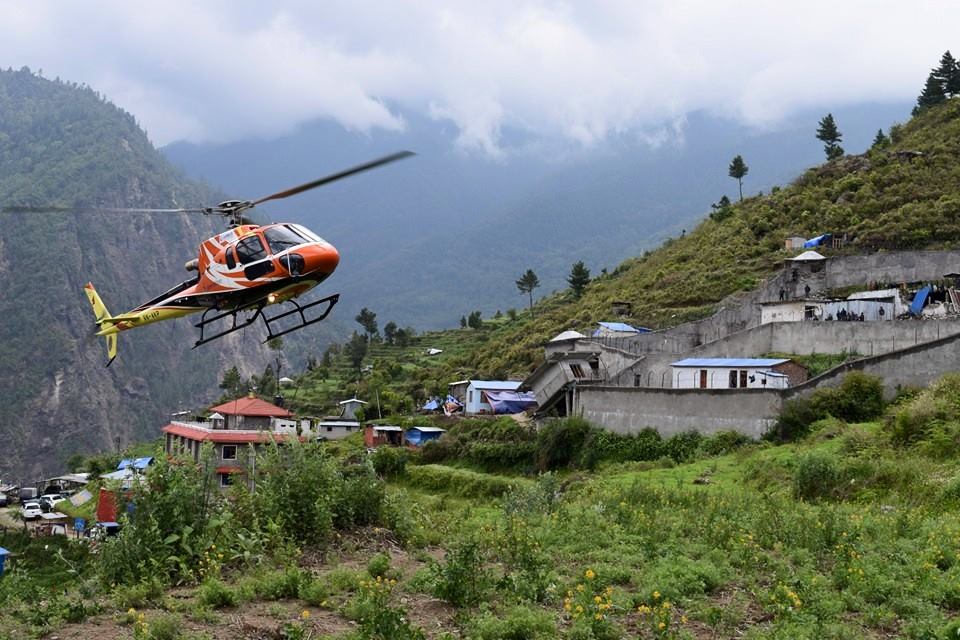}\\

\raisebox{3.2\normalbaselineskip}[0pt][0pt]{\rotatebox[origin=c]{90}{\textbf{Ecuador}}}
&\raisebox{3.2\normalbaselineskip}[0pt][0pt]{\rotatebox[origin=c]{90}{\textbf{Earthquake}}}
&\includegraphics[width=0.28\textwidth]{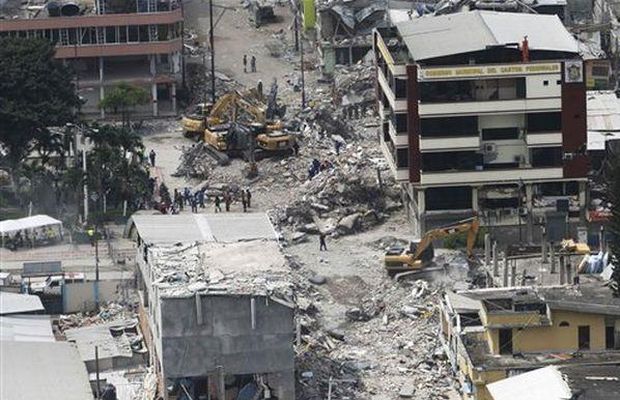}
&\includegraphics[width=0.28\textwidth]{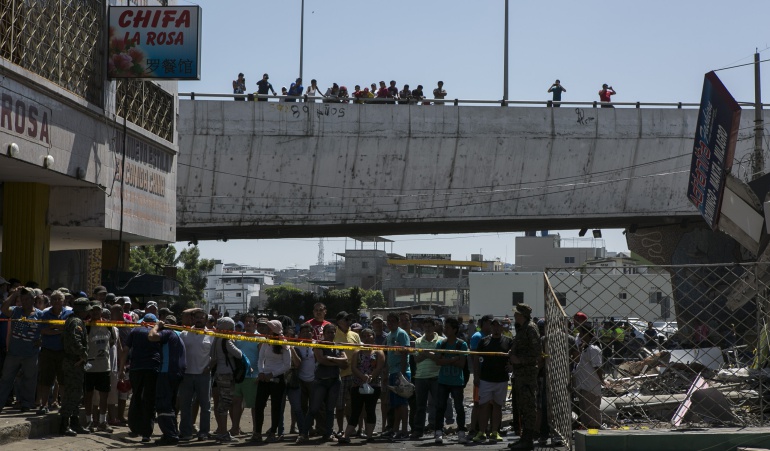}
&\includegraphics[width=0.28\textwidth]{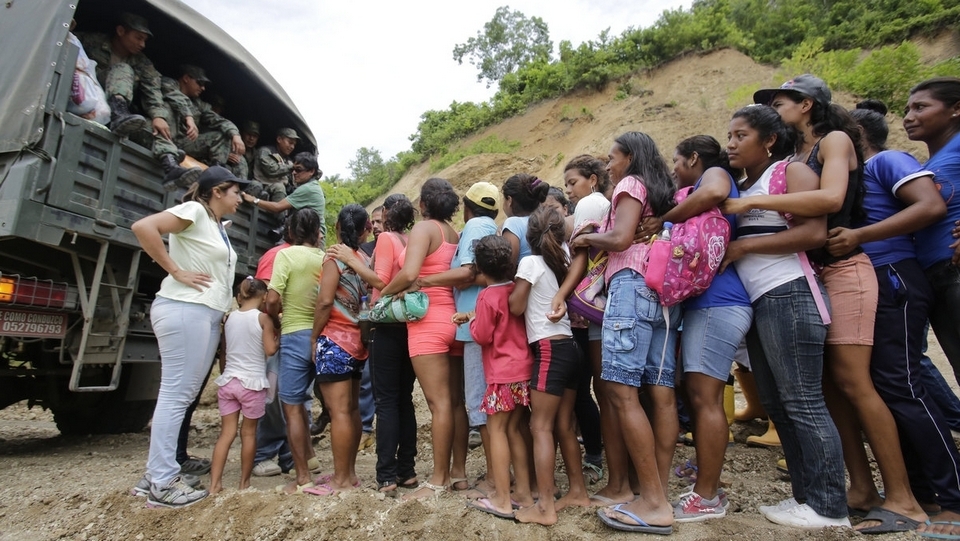}\\

\raisebox{3.2\normalbaselineskip}[0pt][0pt]{\rotatebox[origin=c]{90}{\textbf{Hurricane}}}
&\raisebox{3.2\normalbaselineskip}[0pt][0pt]{\rotatebox[origin=c]{90}{\textbf{Matthew}}}
&\includegraphics[width=0.28\textwidth]{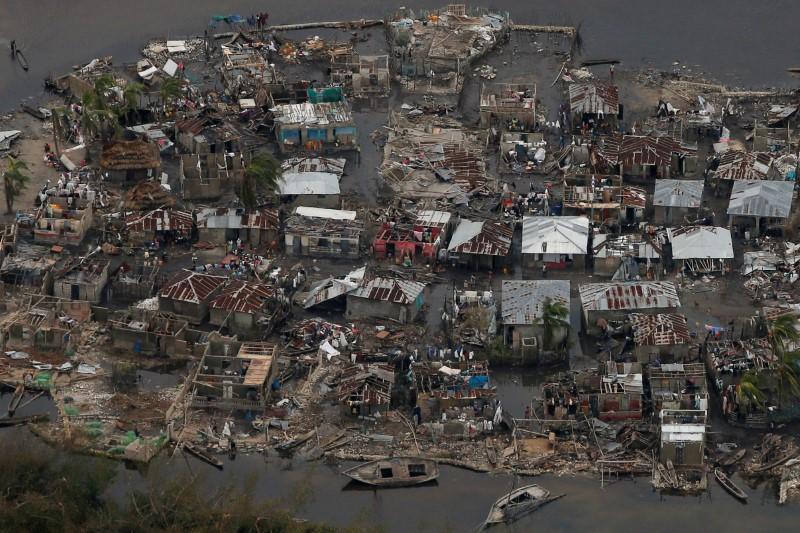}
&\includegraphics[width=0.28\textwidth]{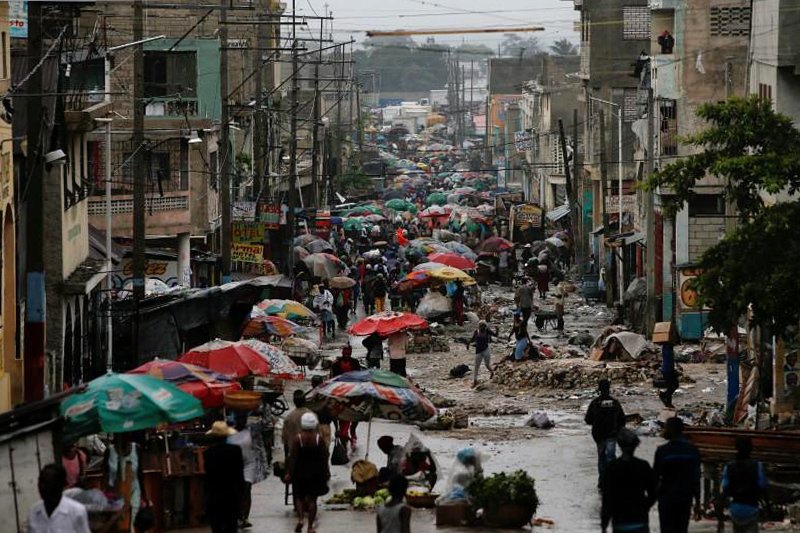}
&\includegraphics[width=0.28\textwidth]{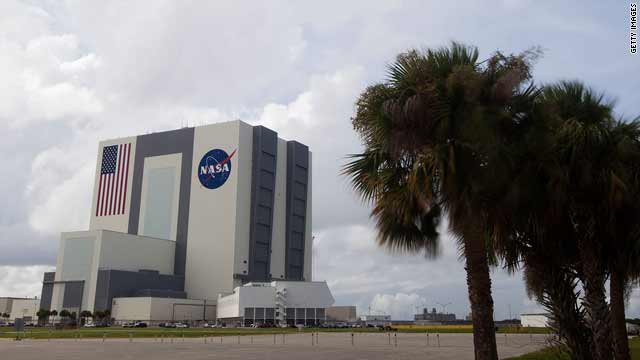}\\

\raisebox{3.2\normalbaselineskip}[0pt][0pt]{\rotatebox[origin=c]{90}{\textbf{Typhoon}}}
&\raisebox{3.2\normalbaselineskip}[0pt][0pt]{\rotatebox[origin=c]{90}{\textbf{Ruby}}}
&\includegraphics[width=0.28\textwidth]{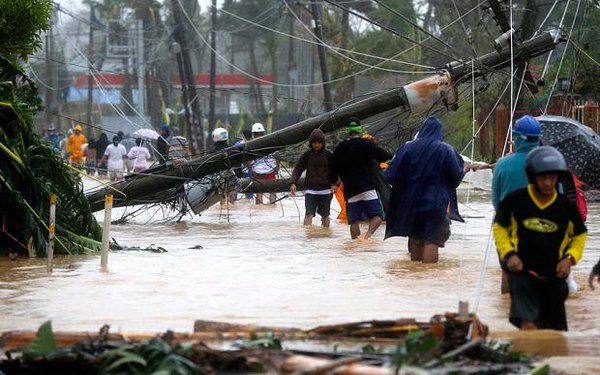}
&\includegraphics[width=0.28\textwidth]{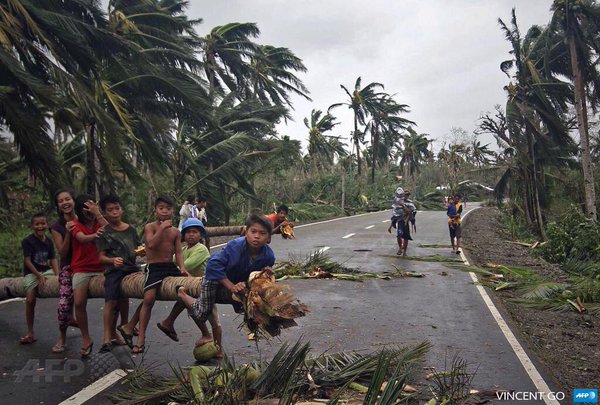}
&\includegraphics[width=0.28\textwidth]{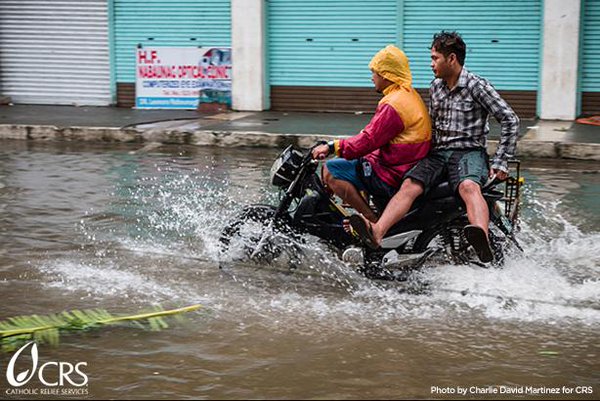}\\
\end{tabular}
\caption{Sample images with different damage levels from different disaster datasets.}
\label{fig:sample_crisis_images}
\end{figure*}

%
We used publicly available AIDR platform~\parencite{imran2014aidr} to collect images from social media networks such as Twitter during four major natural disasters, namely, Typhoon Ruby, Nepal Earthquake, Ecuador Earthquake, and Hurricane Matthew. The data collection was based on event-specific hashtags and keywords.
Table~\ref{tbl:data} lists the total number of images initially collected for each dataset. Figure~\ref{fig:sample_crisis_images} shows example images from these datasets. 




\begin{table}[htb]
\caption{Dataset details for all four disaster events with their year and number of images.}
\centering
\begin{tabular}{rrr}
\toprule
 Disaster Name & Year & Number of Images\\
\midrule
Typhoon Ruby &2014 & 7,000 \\
Nepal Earthquake & 2015 & 57,000\\
Ecuador Earthquake &2016 & 65,000 \\
Hurricane Matthew &2016  & 61,000\\
\bottomrule
\end{tabular}
\label{tbl:data}
\end{table}

\subsection{Human Annotations}
We acquire human labels with the purpose of training and evaluating machine learning models for image filtering and classification. Although, there are several other uses of images from Twitter, in this work, we focus on damage severity assessment from images. Damage assessment is one of the critical situational awareness tasks for many humanitarian organizations. For this purpose, we obtain labels in two different settings. The first set of labels were gathered from AIDR. In this case, volunteers\footnote{Stand-By-Task-Force}, are employed to label images during a crisis situation. In the second setting, we use the Crowdflower\footnote{http://crowdflower.com/}, which is a paid crowdsourcing platform, to annotate images. We provided the following set of instructions with example images for each category to the annotators for the image labeling task. To maintain high-quality, we required an agreement of at least three different annotators to finalize a task.

\subsubsection{Damage Severity Levels Instructions}
\textit{The purpose of this task is to assess the severity of damage shown in an image. The severity of damage in an image is the extent of physical destruction shown in it. We are only interested in physical damages like broken bridges, collapsed or shattered buildings, destroyed or cracked roads, etc. An example of a non-physical damage is the signs of smoke due to fire on a building or bridge---in this particular task, we do not consider such damage types. So in such cases, please select the ``no damage'' category.}

\noindent\textit{\textbf{1- Severe damage}: Images that show the substantial destruction of an infrastructure belong to the severe damage category. A non-livable or non-usable building, a non-crossable bridge, or a non-driveable road are all examples of severely damaged infrastructures.}


\noindent\textit{\textbf{2- Mild damage}:
Damage is generally exceeding minor [damage] with up to 50\% of a building, for example, in the focus of the image sustaining a partial loss of amenity/roof.  Maybe only part of the building has to be closed down, but other parts can still be used. In the case of a bridge, if the bridge can still be used, however, part of it is unusable and/or needs some amount of repairs. Moreover, in the case of a road image, if the road is still usable, however, part of it has to be blocked off because of damage. This damage should be substantially more than what we see due to regular wear or tear.}


\noindent\textit{\textbf{3- Little-to-no damage}:
Images that show damage-free infrastructure (except for wear and tear due to age or disrepair) belong to this category.}

Table~\ref{tbl:labels} shows human annotation results in terms of labeled images for each dataset. Since the annotation was done on raw image collection without any type of prior filtering to clean the dataset, the resulting labeled dataset contains duplicate and irrelevant images. 


\begin{table}[htb]
\caption{Number of labeled images for each dataset and each damage category.}
\centering
\begin{tabular}{rrrrrr}
\toprule
Category&Nepal Earthquake&Ecuador Earthquake&Typhoon Ruby&Hurricane Matthew&Total\\
\midrule
 Severe & 8,927 &  955 & 88 & 110 & 10,080 \\
 Mild & 2,257 &  89  & 338 & 94 & 2,778 \\
 None & 14,239 &  946  & 6,387 & 132 & 21,704 \\
\midrule
 Total  & 25,423 & 1,990 & 6,813 & 336 & 34,562 \\
\bottomrule
\end{tabular}
\label{tbl:labels}
\end{table}

\section{Real-time Filtering of Images}

\begin{figure}[!htb]
\centering \includegraphics[width=\textwidth]{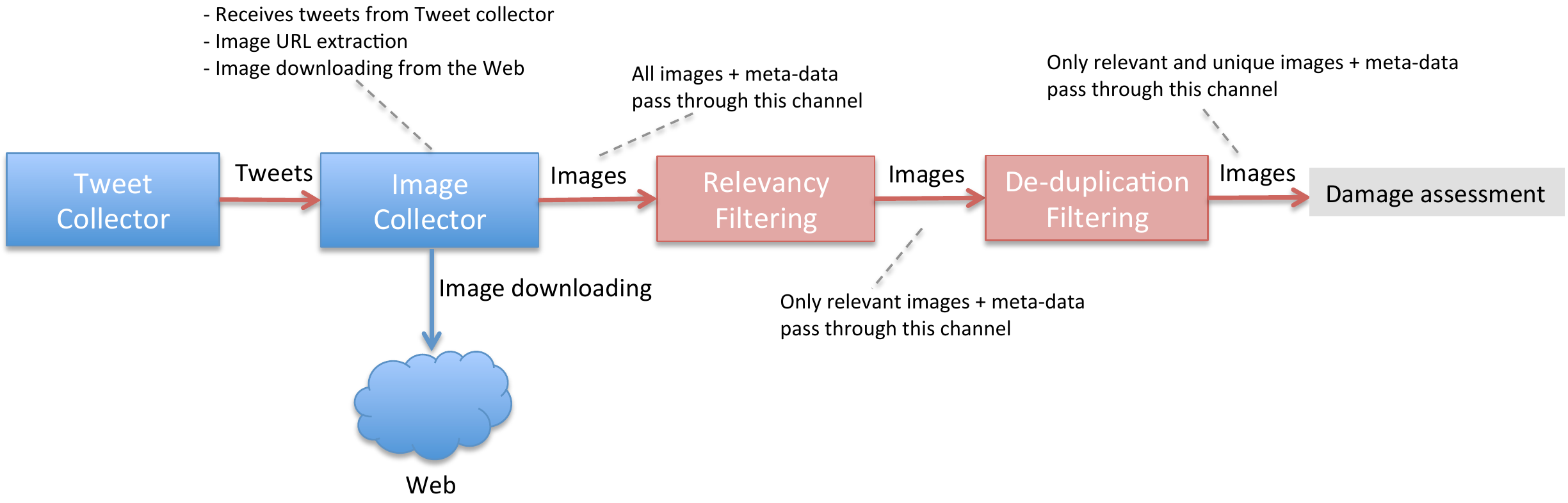}
\caption{Automatic image filtering pipeline.
}
\label{fig:pipeline}
\end{figure}

To be effective during disasters, humanitarian organizations require real-time insights from the data posted on social networks at the onset of any emergency event. To fulfill such time-critical information needs, the data should be processed as soon as it arrives. That means the system should ingest data from online platforms as it is being posted, perform processing and analysis to gain insights in near real-time. To achieve these capabilities, in this paper, we present an automatic image filtering pipeline. Figure~\ref{fig:pipeline} shows the pipeline and its various important components.

The first component in the pipeline is \emph{Tweet Collector}, which is responsible for collecting tweets from Twitter streaming API during a disaster event. A user can collect tweets using keywords, hashtags, or geographical bounding boxes. Although, the pipeline can be extended to consume images from other social media platforms such as Facebook, Instagram, etc., in this paper we only focus on collecting images that are shared via the Twitter platform. The \emph{Image Collector} component receives images from the Tweet collector and extracts image URLs from tweets. Next, given the extracted URLs, the Image Collector downloads images from the Web (i.e., in many cases from Flicker or Instagram). Then, there are the two most important components of the pipeline, i.e., the \emph{Relevancy Filtering} and \emph{De-duplication Filtering}, which we describe next in detail in respective subsections.

\subsection{Relevancy Filtering}

\begin{figure*}[ht!]
\centering
\begin{tabular}{c c c c }
\includegraphics[width=3.5cm,height=3cm]{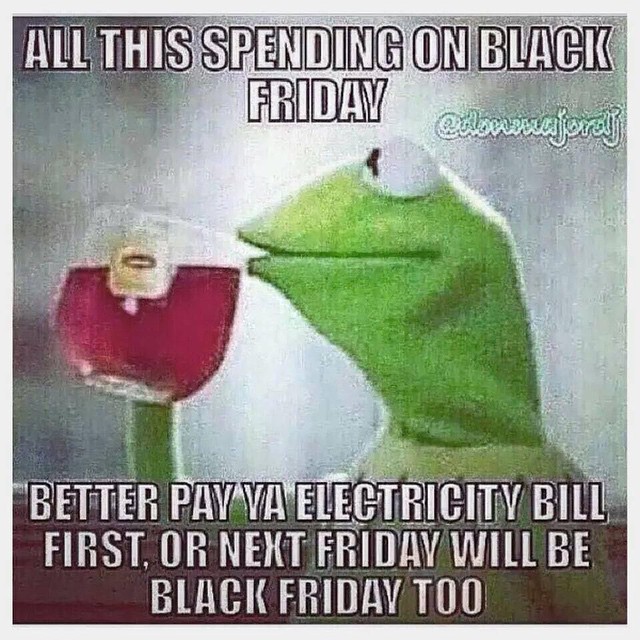} 
&\includegraphics[width=3.5cm,height=3cm]{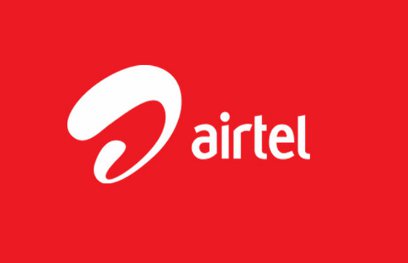}
&\includegraphics[width=3.5cm,height=3cm]{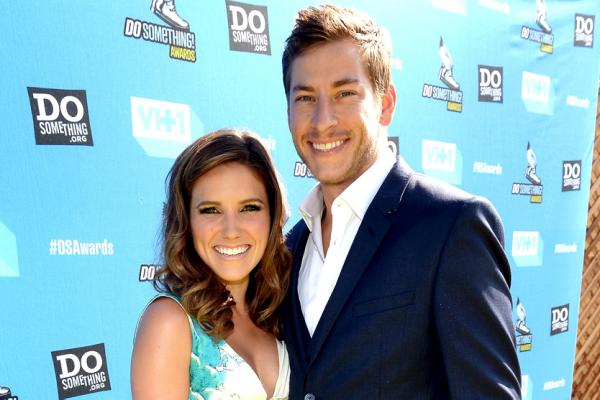}
&\includegraphics[width=3.5cm,height=3cm]{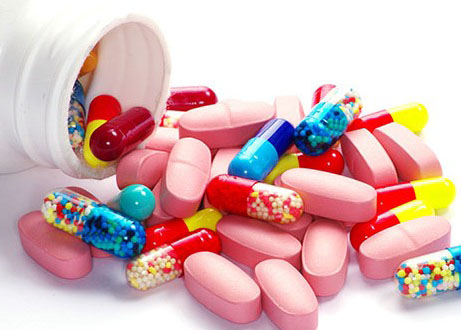}\\
\end{tabular}
\caption{Examples of irrelevant images in our datasets showing cartoons, banners, advertisements, celebrities, etc.}
\label{fig:sample_irrelevant}
\end{figure*}

The concept of \emph{relevancy} depends strongly on the definition and requirements of the task at hand. For example, if the goal is to identify all the news agencies reporting a disaster event, then images that display logos or banners constitute the most relevant portion of the data. On the other hand, if the goal is to identify the level of infrastructure damage after a disaster event, then images that contain buildings, roads, bridges, etc., become the most relevant and informative data samples. Since we have chosen damage assessment as our usecase, and collected human annotations for our datasets accordingly, our perception of relevancy in the scope of this study is strongly aligned with the latter example. Figure~\ref{fig:sample_irrelevant} illustrates example images that are potentially not relevant for damage assessment task.

It is important to note however that the human annotation process presented in the previous section is designed mainly for assessing the level of damage observed in an image, but no question is asked regarding the relevancy of the actual image content. Hence, we lack ground truth human annotations for assessing specifically the relevancy of an image content. We could construct a set of rules or hand-design a set of features to decide whether an image is relevant or not, but we have also avoided such an approach in order not to create a biased or restricted definition of relevancy that may lead to discarding potentially relevant and useful data.
Instead, we have decided to rely on the human-labeled data to learn a set of image features that represent the subset of irrelevant images in our datasets, following a number of steps explained in the sequel.

\subsubsection{Understanding the Content of Irrelevant Images}
We assume the set of images labeled by humans as severe or mild belong to the relevant category. The set of images labeled as none, however, may contain two types of images (i) that are still related to disaster event but do not simply show any damage content, and (ii) that are not related to the disaster event at all, or the relation cannot be immediately understood just from the image content. So, we first try to understand the kind of content shared by the latter set of images, i.e., images that were originally labeled as none. For this purpose, we take advantage of the recent advances in computer vision domain, thanks to the outstanding performance of the convolutional neural networks (CNN), particularly in the object classification task. In this study, we use VGG-16~\parencite{simonyan2014very} as our reference model, which is one of the state-of-the-art deep learning object classification models that performed the best in identifying 1000 object categories in ILSVRC 2014\footnote{http://image-net.org/challenges/LSVRC/2014/results\#clsloc}. To understand which ImageNet object categories appear the most, we first run all of our images labeled by humans as none through the VGG-16 network to classify each image with an object category. We then look at the distribution of the most frequently occurring object categories, and include in our irrelevant set those images that are tagged as one of the following most-frequently-occurring 12 categories: \emph{website, suit, lab coat, envelope, dust jacket, candle, menu, vestment, monitor, street sign, puzzle, television, cash machine, screen}. Finally, this approach yields a set of 3,518 irrelevant images. We also sample a similar number of images randomly from our relevant image set (i.e., images that are originally labeled as severe or mild) to create a balanced dataset of 7,036 images for training the desired relevancy filter as a binary classifier.

\subsubsection{Building a Binary Classifier by Fine-tuning a Pre-trained CNN}
The features learned by CNNs are proven to be transferable and effective when used in other visual recognition tasks~\parencite{yosinski2014transferable,ozbulak2016transferable}, particularly when training data are limited and learning a successful CNN from scratch is not feasible due to overfitting. Considering that we also have limited training data, we adopt a transfer learning approach, where we use the existing weights of the pre-trained VGG-16 network (i.e., millions of parameters, trained on millions of images from the ImageNet dataset) as an initialization for fine-tuning the same network on our own training dataset. We also adapt the last layer of the network to handle binary classification task (i.e., two categories in the softmax layer) instead of the original 1,000-class classification. Hence, this transfer learning approach allows us to transfer the features and the parameters of the network from the broad domain (i.e., large-scale image classification) to the specific one (i.e., relevant-image classification). The details of the training process and the performance of the trained model is reported in the Experimental Framework section.



\subsection{De-duplication Filtering}
\begin{figure*}[ht!]
\centering
\begin{tabular}{c c c c }
  		Blurred and unblurred & With and without text & Cropped  & Re-sized  \\
\includegraphics[width=0.2\textwidth]{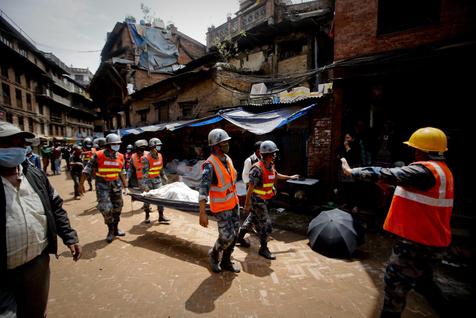} 
&\includegraphics[width=0.2\textwidth]{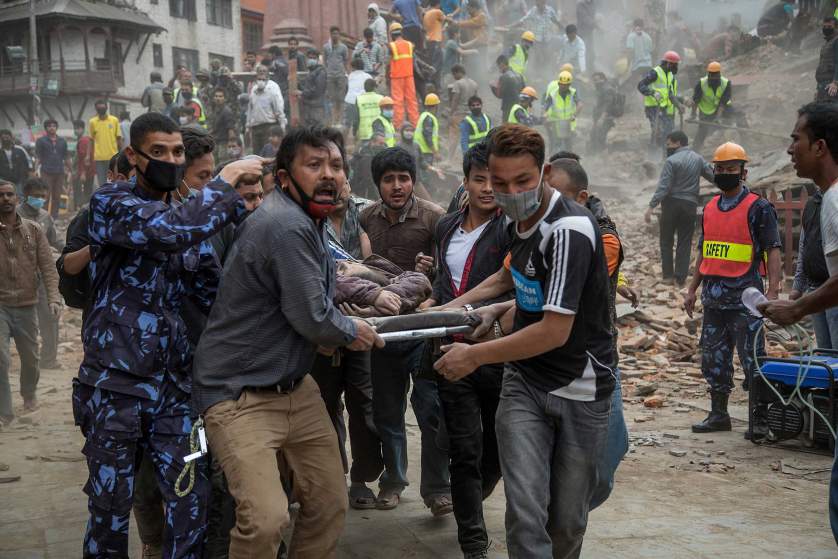}
&\includegraphics[width=0.2\textwidth]{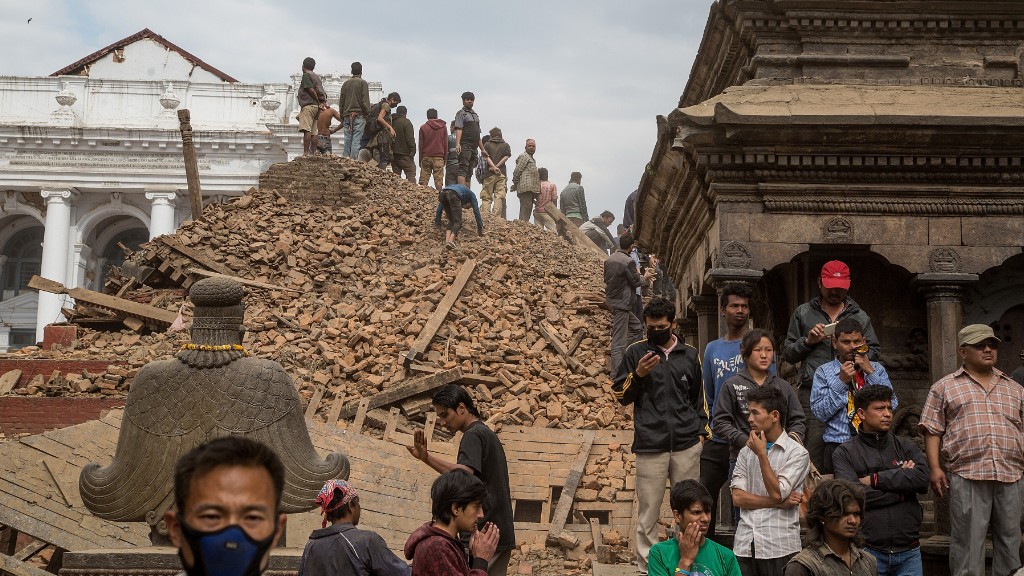}
&\includegraphics[width=0.2\textwidth]{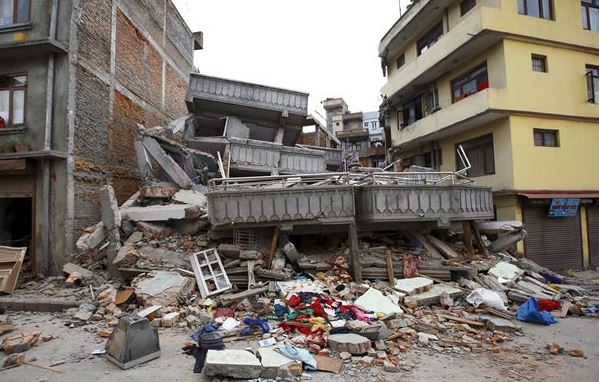}\\
\includegraphics[width=0.2\textwidth]{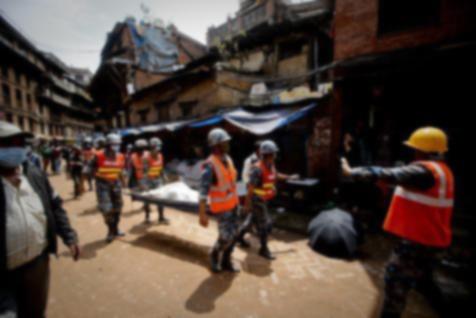} 
&\includegraphics[width=0.2\textwidth]{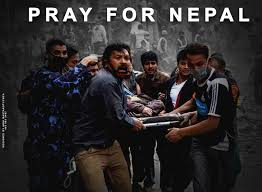}
&\includegraphics[width=0.2\textwidth]{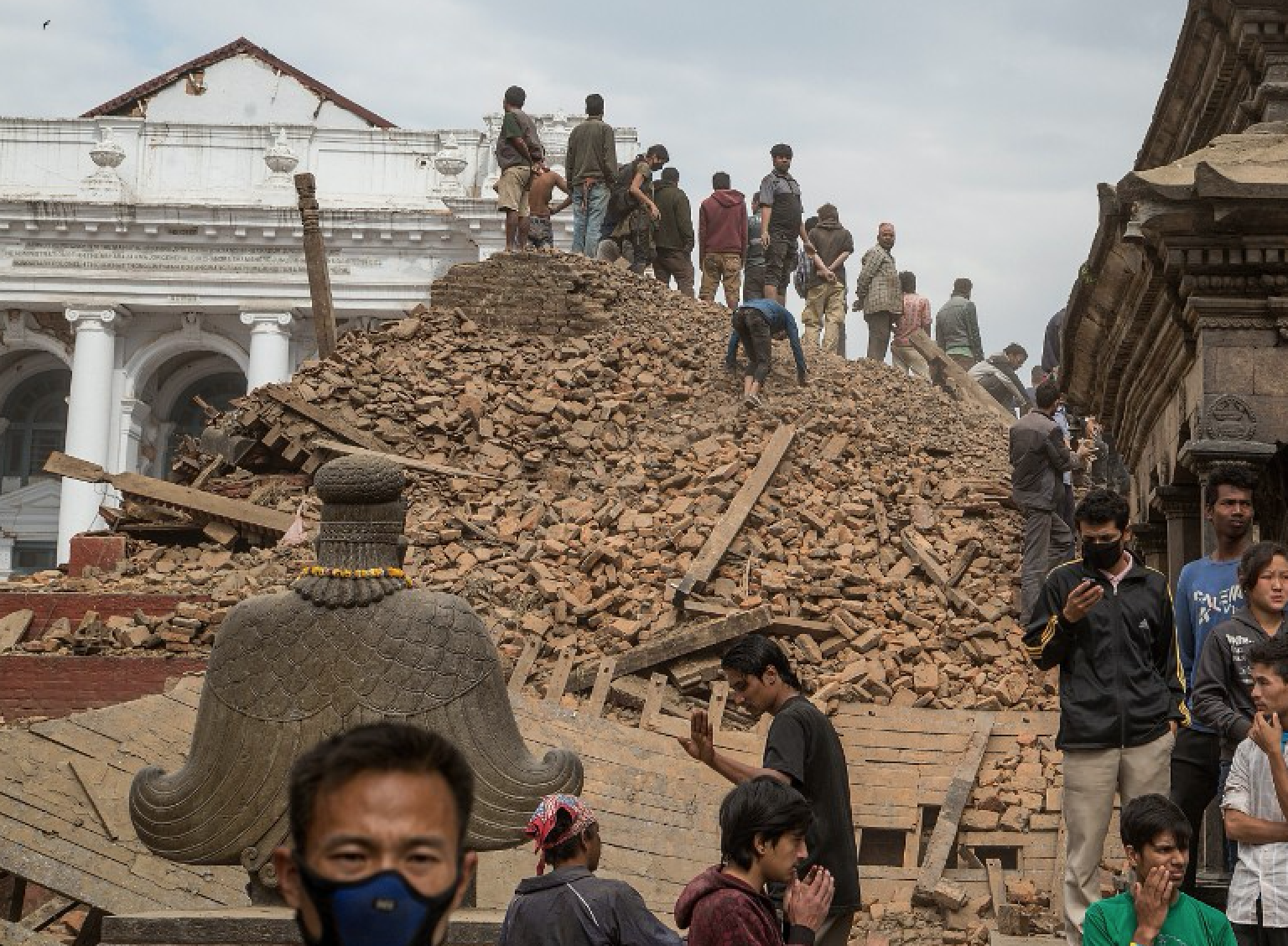}
&\includegraphics[width=0.15\textwidth]{imgs/resize1.jpg}\\
\end{tabular}
\caption{Examples of near-duplicate images found in our datasets.}
\label{fig:sample_duplicate}
\end{figure*}

A large proportion of the image data posted during disaster events contains duplicate or near-duplicate images. For example, there are cases when people simply re-tweet an existing tweet containing an image, or they post images with little modification (e.g., cropping/resizing, background padding, changing intensity, embedding text, etc.). Such posting behavior produces a high number of near-duplicate images in an online data collection. Figure~\ref{fig:sample_duplicate} shows some examples of near-duplicate images.

In order to train a supervised machine learning classifier, as in our case, human-annotation has to be performed on a handful of images. In this case, ideally, train and test sets should be completely distinct from each other. However, the presence of duplicate images may violate this fundamental machine learning assumption, that is, the training and test sets become polluted as they may contain similar images. This phenomenon introduces a bias which leads to an artificial increase in the classification accuracy. Such a system, however, operates at a higher misclassification rate when predicting unseen items.



To detect exact as well as near-duplicate images, we use the Perceptual Hashing (pHash) technique~\parencite{lei2011robust, zauner2010implementation}. Unlike the cryptographic hash functions like MD5~\parencite{rivest1992md5} or SHA1, a perceptual hash represents the fingerprint of an image derived from various features from its content. An image can have different digital representation, for example, due to cropping, resizing, compression or histogram equalization. Traditional cryptographic hashing techniques are not suitable to capture such changes in the binary representation, which is a common case in near duplicate images. Whereas perceptual hash functions maintain \emph{perceptual equality} of images hence they are robust in detecting even slight changes in the binary representation of two similar images.

The de-duplication filtering component in our system implements the Perceptual Hashing technique\footnote{\url{http://www.phash.org/}} to determine whether or not a given pair of images are same or similar. Specifically, it extracts certain features from each image, and computes a hash value for each image based on these features, and compares the resulting pair of hashes to decide the level of similarity between the images. During an event, the system maintains a list of hashes computed for a set of distinct images it receives from the Image collector module. To determine whether a newly arrived image is a duplicate of an already existing image, hash value of the new image is computed and compared against the list of stored hashes to calculate its distance from the existing image hashes. In our case, we use the Hamming distance to compare two hashes. If an image with a distance value smaller than $d$ is found in the list, the newly arrived image is considered as duplicate. The details regarding the optimal value of $d$ are given in the Experimental Framework section. We always keep the recent 100k hashes in our physical memory for fast comparisons. This number obviously depends on the size of available memory in the system.

\section{Experimental Framework}

For the performance evaluation of the different system components, we report on several well-known metrics such as accuracy, precision, recall, F1-score and AUC. Accuracy is computed as the proportion of correct predictions, both positive and negative. Precision is the fraction of the number of true positive predictions to the number of all positive predictions. Recall is the fraction of the number of true positive predictions to the actual number of positive instances. F1-score is the harmonic mean of precision and recall. AUC is computed as the area under the precision-recall curve.

\subsection{Tuning the Individual Image Filtering Components}

In this section, we elaborate on the tuning details of our individual relevancy and de-duplication filtering modules as well as their effects on raw data collection from online social networks.

\subsubsection{Training and Testing the Performance of the Relevancy Filtering}
We use 60\% of our 7,036 images for training and 20\% for validation during fine-tuning of the VGG-16 network. We then test the performance of the fine-tuned network on the remaining 20\% of the dataset. Table~\ref{tbl:relevancy_performance} presents the performance of the resulting relevancy filter on the test set. Almost perfect performance of the binary classifier stems from the fact that relevant and irrelevant images in our training dataset have completely different image characteristics and content (as can be seen from the example images in Figures~\ref{fig:sample_crisis_images} and~\ref{fig:sample_irrelevant}).
This meets our original relevancy filtering objective to remove only those images that are surely irrelevant to the task at hand. Note that we reserve these 7,036 images only for relevancy filter modeling, and perform the rest of the experiments presented later using the remaining 27,526 images.

\begin{table}[htb]
\caption{Performance of the relevancy filter on the test set.}
\centering
\begin{tabular}{cccc}
\toprule
AUC & Precision & Recall & F1 \\
\midrule
0.98 & 0.99 & 0.97  & 0.98 \\
\bottomrule
\end{tabular}
\label{tbl:relevancy_performance}
\end{table}


\subsubsection{Learning an Optimal Distance Threshold for the De-duplication Filtering}
\begin{figure}[!htb]
\centering
\begin{tikzpicture}[y=4.0cm, x=.34cm,font=\sffamily]
	\draw (0,0) -- coordinate (x axis mid) (20,0);
    	\draw (0,0) -- coordinate (y axis mid) (0,1);
    	\foreach \x in {0,2,...,20}
     		\draw (\x,1pt) -- (\x,-3pt)
			node[anchor=north] {\x};
    	\foreach \y in {0,0.2,0.4,0.6,0.8,1}
     		\draw (1pt,\y) -- (-3pt,\y) 
     			node[anchor=east] {\y}; 
	\node[below=0.8cm] at (x axis mid) {Hamming Distance ($d$)};
	\node[rotate=90, above=0.8cm] at (y axis mid) {Accuracy};
	\draw plot[mark=*, mark options={fill=white}] 
		file {syn2.data};	  
\end{tikzpicture}
\caption{Estimation of distance threshold $d$ for duplicate image detection.}
\label{fig:duplicate}
\end{figure}
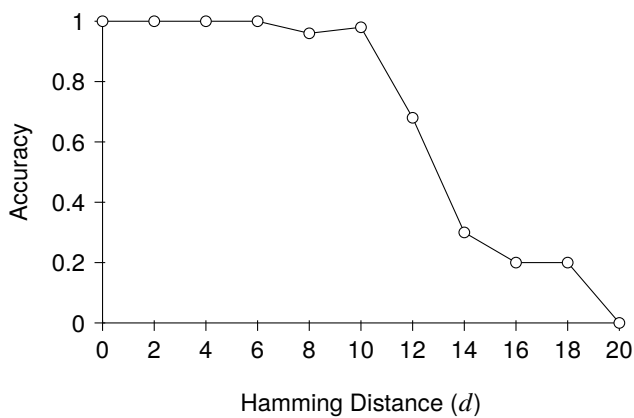

To detect duplicate images, we first learned an optimal distance $d$ as our threshold to decide whether two images are similar or distinct, (i.e., two images with distance $\leq d$ are considered duplicate). For this purpose, we randomly selected 1,100 images from our datasets and computed their pHash values. Next, Hamming distance for each pair is computed. To determine the optimal distance threshold, we manually investigated all pairs with a distance between 0 to 20. Pairs with distance $> 20$ look very distinct, thus not selected for manual annotation. We examined each pair and assigned a value of 1, if the images in a pair are the same and 0 otherwise. Figure~\ref{fig:duplicate} shows the accuracy computed based on the human-annotations, i.e., the number of correct predictions (including duplicate and not-duplicate) by machine at a given threshold value. We can clearly observe high accuracy with distance values $d\leq10$, however, after that the accuracy decreases drastically. Based on this experiment, a distance threshold $d=10$ is selected.

\subsubsection{Effects of the Individual Image Filtering Components on Raw Data Collection}
To illustrate the benefit of image filtering components as well as to understand the useful proportion of the incoming raw image data from online social networks, we apply our proposed relevancy and de-duplication filtering modules on the remaining 27,526 images in our dataset. Table~\ref{tbl:data_after_filtering} presents the number of images retained in the dataset after each image filtering operation. As expected, relevancy filtering eliminates 8,886 of 18,186 images in none category, corresponding to an almost 50\% reduction. There are some images removed from the severe and mild categories (i.e., 212 and 164, respectively) but these numbers are in the acceptable range of 2\% error margin for the trained relevancy classifier as reported earlier. De-duplication filter, on the other hand, removes a considerable proportion of images from all categories, i.e., 58\%, 50\% and 30\% from severe, mild and none categories, respectively. The relatively higher removal rate for the severe and mild categories can be explained by the fact that social media users tend to re-post the most relevant content more often. Consequently, our image filtering pipeline reduces the size of the raw image data collection by almost a factor of 3 (i.e., an overall reduction of 62\%) while retaining the most relevant and informative image content for further analyses which we present in the next section.

\begin{table}[htb]
\caption{Number of images that remain in our dataset after each image filtering operation.
}
\centering
\begin{tabular}{rrrrr}
\toprule
Category&Raw Collection&After Relevancy Filtering&After De-duplication Filtering&Overall Reduction\\
\midrule
 Severe & 7,501 & 7,289 & 3,084 & 59\%\\
 Mild   & 1,839 & 1,675 & 844 & 54\%\\
 None   & 18,186 & 9,300 & 6,553 & 64\%\\
\midrule
All & 27,526 & 18,264 & 10,481 & \textbf{62\%}\\
\bottomrule
\end{tabular}
\label{tbl:data_after_filtering}
\end{table}

\subsection{Evaluating the Effects of Image Filtering Components on Damage Assessment Performance}

In this section, we analyze the effects of irrelevant and duplicate images on human-computation and machine training. We choose the task of damage assessment from images as our usecase, and consider the following four settings:

\begin{itemize}
\setlength\itemsep{0em}
\item S1: We perform experiments on raw collection by keeping duplicate and irrelevant images intact.
The results obtained from this setting are considered as a baseline for the next settings.

\item S2: We only remove duplicate images and keep the rest of the data the same as in S1. The aim of this setting is to learn the difference with and without duplicates.

\item S3: We only remove irrelevant images and keep the rest of the data the same as in S1.
The aim here is to investigate the effects of removing irrelevant images from the training set.

\item S4: We remove both duplicate and irrelevant images. This is the ideal setting, which is also implemented in our proposed pipeline. This setting is expected to outperform others both in terms of budget utilization and machine performance.
\end{itemize}

For training a damage assessment classifier, we again take the approach of fine-tuning a pre-trained VGG-16 network. This is the same approach that we use for relevancy filter modeling before, but this time (i) the network is trained for 3-class classification where classes are severe, mild, and none, and (ii) the performance of the resulting damage assessment classifier is evaluated in a 5-fold cross-validation setting rather than using a train/validation/test data split.

\subsubsection{Effects of Duplicate Images on Human Annotations and Machine Training}
To mimic the real-time annotation and machine learning setting during a crisis event, we assume that a fixed budget of 6,000 USD is available for annotation. For simplicity, we assume 1 USD is the cost to get one image labeled by human workers. Given this budget, we aim to train two classifiers; one with duplicates (S1) and another without duplicates (S2).

In S1, the system spends full 6,000 USD budget to get 6,000 labeled images from the raw collection, many of which are potential duplicates. 
To simulate this, we randomly select 6,000 images from our labeled dataset while maintaining the original class distributions as shown in the S1 column of Table~\ref{tbl:im2train}. We then use these 6,000 images to train a damage assessment classifier as described above, and present the performance of the classifier in the S1 column of Table~\ref{tbl:classification_report}. 

In S2, we take the same subset of 6,000 images used in S1, and run them through our de-duplication filter to eliminate potential duplicates, and then, train a damage assessment classifier on the cleaned subset of images. S2 column of Table~\ref{tbl:im2train} shows the class-wise distribution of the remaining images after de-duplication. We see that 598, 121, and 459 images are marked as duplicate and discarded in severe, mild, and none categories, respectively. Evidently, this indicates a budget waste of 1,178 USD ($\sim$20\%) in S1, which could have been saved if the de-duplication technique was employed.
The performance of the damage assessment classifier trained on the cleaned data is shown in S2 column of Table~\ref{tbl:classification_report}. Although we observe an overall decrease in all performance scores as compared to S1, we claim that the performance results for S2 are more trustworthy for the following reason: In S1, due to the appearance of duplicate or near-duplicate images both in training and test sets, the classifier gets biased and thus shows artificially high but unreliable performance.

\begin{table}[htb]
\caption{Number of images used in each setting: S1 (with duplicates + with irrelevant), S2 (without duplicates + with irrelevant), S3 (with duplicates + without irrelevant), S4 (without duplicates + without irrelevant).}
\centering
\begin{tabular}{rrrrr}
\toprule
Category&S1&S2&S3&S4\\
\midrule
 Severe & 1,636 &  1,038 & 2,395 & 1,765 \\
 Mild   & 400 &  279 &  550& 483 \\
 None   & 3,964 &  3,505  & 3,055 & 3,751 \\
\midrule
All & 6,000 & 4,822 & 6,000 & 6,000\\
\bottomrule
\end{tabular}
\label{tbl:im2train}
\end{table}

\begin{table}[htb]
\caption{Precision, Recall, F1 and AUC scores: S1 (with duplicates + with irrelevant), S2 (without duplicates + with irrelevant), S3 (with duplicates + without irrelevant), S4 (without duplicates + without irrelevant).}
\centering
\resizebox{\columnwidth}{!}{%
\begin{tabular}{l c c c c c c c c c c c c c c c c}
\toprule
& \multicolumn{4}{c}{S1} & \multicolumn{4}{c}{S2} & \multicolumn{4}{c}{S3} & \multicolumn{4}{c}{S4} \\ \cmidrule(lr){2-5} \cmidrule(lr){6-9} \cmidrule(lr){10-13} \cmidrule(lr){14-17}

	& AUC & Pre. & Rec. &F1 & AUC & Pre. & Rec. &F1 & AUC & Pre. & Rec. &F1 & AUC & Pre. & Rec. &F1 \\
\midrule
None  & 0.98 & 0.91 & 0.96 & 0.94 &0.98& 0.91 & 0.97 & 0.94 & 0.94& 0.86 & 0.93 & 0.90 &0.95& 0.86 & 0.95 & 0.91\\
Mild  & 0.31 &0.48 & 0.18 & 0.25  &0.26& 0.41 & 0.12 & 0.18 & 0.37& 0.53 & 0.20 & 0.29 &0.30& 0.55 & 0.14 & 0.23\\
Severe& 0.95 & 0.88 & 0.89 & 0.88 &0.91& 0.85 & 0.84 & 0.84 & 0.95 & 0.88 & 0.91 & 0.90 &0.91& 0.86 & 0.85 & 0.86\\
 
\midrule
Avg. & 0.75 & 0.74 & 0.68 & 0.69 & 0.72 & 0.72 & 0.64 & 0.65 & 0.75  & 0.76 & 0.68 & 0.70 & 0.72 & 0.76 & 0.65 & 0.67\\
\bottomrule
\end{tabular}}
\label{tbl:classification_report}
\end{table}

\subsubsection{Effects of Irrelevant Images on the Damage Assessment Performance}
In terms of machine training, the level of noise in a dataset determines the quality of the trained model, and hence, the overall performance of the designed system (i.e., high level of noise leads to a low-quality model, which in turn, leads to a poor system performance). In terms of human annotation, a noisy dataset also causes sub-optimal use of the available budget. Even though removing irrelevant images from our datasets helps us in both directions, in this section we focus only on the machine training aspect of the problem as we have already exemplified the effect of such an operation on human annotation in the previous section.

Recall that in S1 we sample 6,000 images directly from the raw collection of 27,526 images. In S3 we first apply relevancy filtering operation on the raw collection and then sample 6,000 images from the clean set of 18,264 images. Note that this 6,000 sample may still contain duplicate or near-duplicate images. Even though the training data for S1 and S3 are not exactly the same, we can still try to compare the performance of the damage assessment model with and without irrelevant images. As we see from S1 and S3 columns of Table~\ref{tbl:classification_report}, the scores for none category in S3 are lower than those in S1 whereas the scores for severe and mild categories in S3 are higher than those in S1. After macro-averaging the scores for all categories, we see that overall F1-score for S3 is 1\% higher than the overall F1-score for S2. Though macro-averaged AUC scores seem to be the same. However, we already know from the previous section that having duplicate or near-duplicate images in the training set yield untrustworthy model results. Hence, we do not intend to elaborate any further on this comparison.

\begin{figure}[!htb]
\centering
\begin{tabular}{cc}
S1 & S2 \\
\includegraphics[width=6cm]{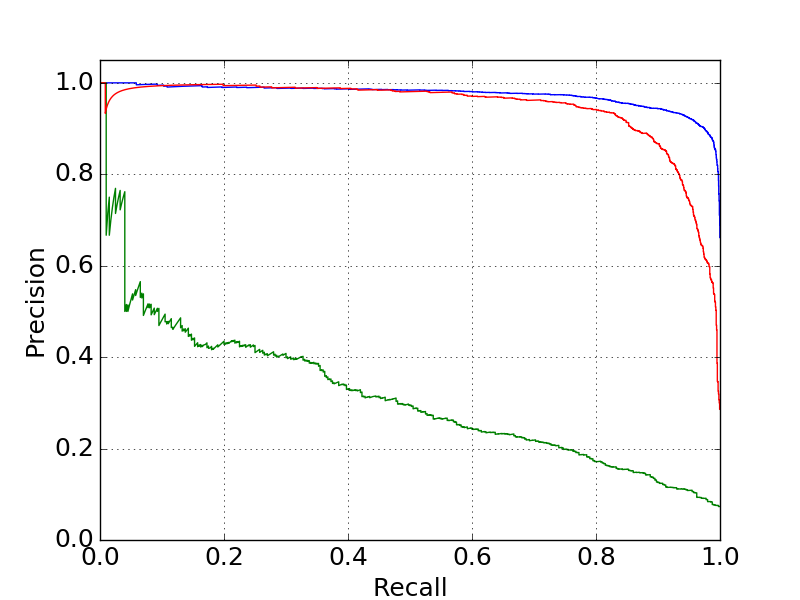} &
\includegraphics[width=6cm]{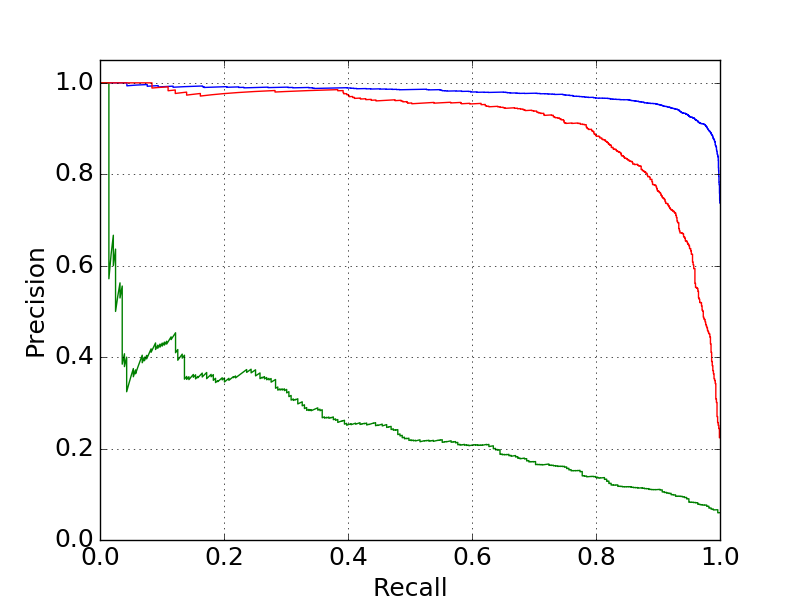}\\\\
S3 & S4  \\
\includegraphics[width=6cm]{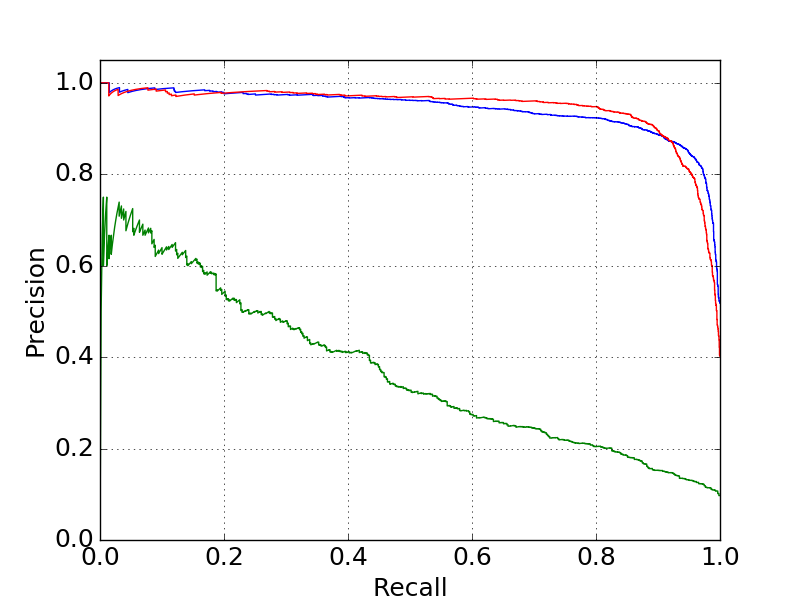} &\includegraphics[width=6cm]{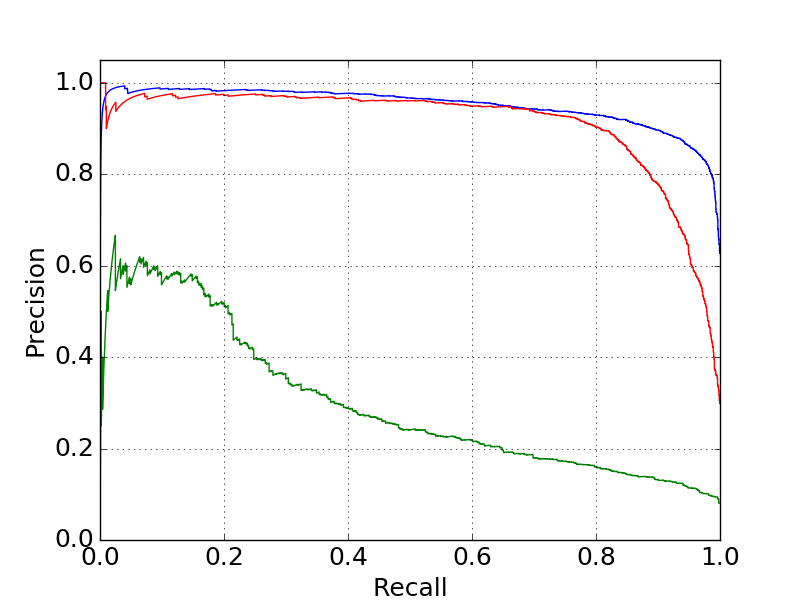} \\
\multicolumn{2}{c}{\includegraphics[width=4cm]{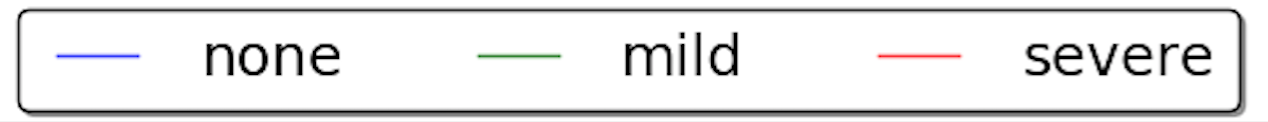}}\\
\end{tabular}
\caption{Precision-recall curves for all four settings: S1 (with duplicates + with irrelevant), S2 (without duplicates + with irrelevant), S3 (with duplicates + without irrelevant), S4 (without duplicates + without irrelevant).}
\label{fig:auc}
\end{figure}

In the ideal setting (S4), we discard both duplicate and irrelevant images from the raw data collection, and then sample 6,000 images from the remaining clean set of 10,481 images. S4 column of Table~\ref{tbl:classification_report} presents the results of the damage assessment experiment on the sampled dataset, which does not contain duplicate and irrelevant images. If we compare the performance results for S3 and S4, we see that removing duplicate images from the training data eliminates the artificial increase in the performance scores, which is in agreement with the trend observed between S1 and S2.

More interestingly, we can see the benefit of removing irrelevant images when the data is already free from duplicates. That is, we can compare the results of S2 and S4, even though the training data for both settings are not exactly the same. At the category level, we observe a similar behavior as before, where the scores for none category in S4 are slightly lower than those in S2 while the scores for severe and mild categories in S4 are slightly higher than those in S2. If we compare the macro-averaged F1-scores, we see that S4 outperforms S2 by a small margin of 2\%. In order to assess whether this difference in F1-scores between S2 and S4 is significant or not, we perform a permutation test (or sometimes called a randomization test) in the following manner. We randomly shuffle 1,000 times the input test image labels and the output model predictions within a common pool of S2 and S4 image subsets. Then for each shuffle, we compute the difference in F1-scores for S2 and S4. Eventually, we compare the observed F1-score distance against the distribution of such sampled 1,000 F1-score differences to see if the observed value is statistically significantly away from the mean of the sample distribution. In our case, we get $p=0.077$, which is not statistically significant but shows a certain trend toward significance.

Figure~\ref{fig:auc} shows the precision-recall curves for all four settings. First of all, it is evident that in the three-class classification task, the hardest class (according to the classifier performance) is the mild damage category. In all settings, we observe a low AUC for mild category compared to the other two categories. One obvious justification of this is the low prevalence of the mild category compared to other categories (see Table~\ref{tbl:im2train}). More training data should fix this issue, which we plan as future work. Otherwise, in all settings, the classifiers achieve high accuracy in classifying images into severe and none categories.

\section{Discussion}
User-generated content on social media at the time of disasters is useful for crisis response and management. However, understanding this high-volume, high-velocity data is a challenging task for humanitarian organizations. In this paper, we presented a social media image filtering pipeline to specifically perform two types of filtering: i) image relevancy filtering and ii) image de-duplication.

This work is a first step towards building more innovative solutions for humanitarian organizations to gain situational awareness and to extract actionable insights in real time. However, a lot has to be done to achieve that goal. For instance, we aim to perform an exploratory study to determine whether there are differences between the key image features for specific event types, e.g., how key features for earthquake images differ from key features for hurricane images. This type of analysis will help us to build more robust classifiers, i.e., either general classifiers for multiple disaster types or classifiers specific to certain disaster types.

Moreover, as in the current work, we use labeled data from two different sources (i.e., paid and volunteers). It is worth performing a quality assessment study of these two types of crowdsourcing, especially in the image annotation context, to understand if there are differences in the quality of label annotation agreements between annotators from two diverse platforms, i.e., Crowdflower \textit{vs.}\ AIDR. We consider this type of a quality assessment study as a potential future work.

Understanding and modeling the \emph{relevancy} of certain image content is another core challenge that needs to be addressed more rigorously. Since different humanitarian organizations have different information needs, the definition of relevancy needs to be adapted or formulated according to the particular needs of each organization. Adapting a baseline relevancy model or building a new from scratch is a decision that has to be made at the onset of a crisis event, which is what we aim to address in our future work. 

Besides relevancy, the veracity of the extracted information is also critical for humanitarian organizations to gain situational awareness and launch relief efforts accordingly. We have not considered evaluating the veracity of images for a specific event in this paper. However, we plan to tackle this important challenge in the future.

\section{Conclusion}

Existing studies indicate the usefulness of imagery data posted on social networks at the time of disasters. However, due to large amounts of redundant and irrelevant images, efficient utilization of the imagery content both using crowdsourcing or machine learning is a great challenge. In this paper, we have proposed an image filtering pipeline, which comprised of filters for detecting irrelevant as well as redundant (i.e., duplicate and near-duplicate) images in the incoming social media data stream. To filter out irrelevant image content, we used a transfer learning approach based on the state-of-the-art deep neural networks. For image de-duplication, we employed perceptual hashing techniques. We also performed extensive experimentation on a number of real-world disaster datasets to show the utility of our proposed image processing pipeline. We hope our real-time online image processing pipeline facilitates consumption of social media imagery content in a timely and effectively manner so that the extracted information can further enable early decision-making and other humanitarian tasks such as gaining situational awareness during an on-going event or assessing the severity of damage during a disaster.

\printbibliography
 \end{document}